# About the Exact Role of Disturbance in Heisenberg's Uncertainty Relation

Johan Wulleman

______________________________________________________________

**Abstract**

It is generally accepted that the disturbance interpretation cannot explain Heisenberg's uncertainty relation $\Delta x \, \Delta p \approx h$. In this paper a clear distinction will be made between the notion of *state preparation* and *measurement*, and the disturbance that is referred to in Heisenberg's disturbance interpretation is usually the disturbance that is caused by the measurement. The main goal of this paper is to examine and discuss to what extent disturbance during the state preparation plays a role in the interpretation of Heisenberg's uncertainty relation. This examination will be done in the context of the single-slit experiment. To conclude this paper a novel single-slit experiment will be proposed and discussed with which a clearer insight could be gained in the exact role of disturbance when a quantum system passes a single-slit.

**Résumé**

Il est généralement accepté que l'interprétation de perturbation ne puisse pas expliquer la relation d'incertitude de Heisenberg $\Delta x \, \Delta p \approx h$. Dans cet article une distinction claire sera faite entre la notion de la *préparation de l'état* et la *mesure*, et la perturbation à laquelle il est fait référence dans l'interprétation de perturbation de Heisenberg qui est généralement la perturbation causée par la mesure. L'objectif principal de cet article est d'étudier et d'examiner dans quelle mesure la perturbation pendant la préparation de l'état joue un rôle dans l'interprétation de la relation d'incertitude de Heisenberg. Cet examen sera fait dans le cadre de l'expérience de fente-unique. Pour conclure cet article, une expérience de fente-unique nouvelle sera proposée et examinée grâce à laquelle un aperçu plus clair pourrait être obtenu du rôle précis de la perturbation quand un système quantique passe une fente-unique.

**Keywords**: Heisenberg's uncertainty relation, disturbance, single-slit experiment, BEC, STM

# 1. Introduction

In 1927 Heisenberg [1] published his famous uncertainty relation

$$\Delta x\, \Delta p \approx h. \qquad (1)$$

One particular interpretation that was shared by Heisenberg himself, namely the disturbance interpretation, is the issue of this paper. Heisenberg's view was that understanding quantum phenomena centers on the idea that a *measurement* disturbs the measured quantum system in an unpredictable way. Heisenberg proposed that the value of an observable is well-defined right after the measurement, and is therefore more or less created during the measurement. According to his view, the measurement causes a disturbance on the value of the observables that are not compatible with the measured one. For Heisenberg it is therefore the measurement itself that determines the final state of the quantum system (or object). This disturbance interpretation, however, has received criticism by several authors, just to name a few; Ballentine [2], Hooker [3] and Brown & Redhead [4].

Ballentine, for instance, discusses the Uncertainty Principle (p. 365) and studies in the context of a single-slit experiment what the role is of the disturbance coming from the measuring devices (array of detectors which make up the screen). His conclusion, which contradicts Heisenberg's conclusion, is that the Uncertainty Principle (UP) "… is not in any real sense related to the possible disturbance of a system by the measurement.", a conclusion that is generally accepted. It is important to note that Ballentine makes a clear distinction between two notions, namely: *state preparation* and the measurement itself; the state preparation being done between the time instance that the system entered the single-slit and the instance just prior to the measurement. As the UP is not related to any disturbance from the measurement, it is then according to Ballentine's conclusion clear that the UP ($\Delta x \Delta p \approx h$) is related to the state preparation. About Ballentine's conclusion I would like to emphasize that the disturbance he mentions comes exclusively from the measurement itself; Ballentine does not study what impact a disturbance coming from the state-preparer (the single-slit) might have on the system (particle) and what role that disturbance could play in the UP during the state preparation.

Hooker's paper then is not directly about Heisenberg's uncertainty relation (or principle) as such but about the notion of disturbance. In his paper (p. 262) Hooker cites Heisenberg's approach of understanding quantum phenomena where the disturbance is considered as the physical interaction between the measured system and the measuring instrument in which a quantum of energy is exchanged during a finite transition time. Hooker then continues and comes to lemma P7 (p. 263), which says that during the finite transition time no definite energy can be assigned to the systems involved in the interaction. Any attempt to attribute a definite energy to P7 leads to a contradiction in the reasoning, which is, according to Hooker, incompatible with the "disturbance" interpretation. Unfortunately, Hooker is in his paper not quite clear how the disturbance interpretation cannot explain the uncertainty relation. In fact, from his study it looks like he puts in question the (his) notion of disturbance itself. But that is not what we see in any collision or Compton experiment! For instance, an incident photon that collides with an object system will effectively interact with the object, exchange energy and momentum with it and finally scatter away, which results in a different energy and momentum than before for both the photon and the object. So, whatever happens during the time of the interaction between the photon and object, whether or not we can assign a definite energy during the interaction time, it is an *experimental fact* that an interaction takes effectively place as well as an energy and momentum exchange. An in-depth discussion about Hookers reasoning is not within the scope of this paper but it should be clear that it is not the validity of the notion of disturbance as such in a physical interaction that should be the issue of the discussion but the role that any disturbance can play in the interpretation of Heisenberg's uncertainty relation.

Brown & Redhead discuss a very particular aspect of uncertainty, namely the *transfer* of uncertainty from one system to another during an interaction. In the case of Heisenberg's microscope this would mean that the uncertainty relation comes from the incident photon's quantal uncertainty (indeterminacy) that has been transferred to the object system under study (and to the scattered photon) during the interaction at the time of scattering. The obvious question Brown & Redhead ask themselves is then where the quantal uncertainty of the incident photon comes from? The idea that the uncertainty relation is coming from another system through the transfer of uncertainty (via an interaction in their case) is indeed untenable. However, the most worrying is maybe in the last paragraph of their conclusion (p. 18), which says that the source of the uncertainty relations is *not* to be sought in the disturbance by interaction with another quantal system. This conclusion shows at least a few weaknesses. First, how well can this conclusion be generalized to the case where the incident quantum system interacts with another "quantal system" that is very massive as in the case of a single-slit device? Secondly, apart from the fact that an interaction can transfer uncertainty, the interaction itself could also be responsible for generating an additional disturbance during the state preparation (as will be illustrated in the next section with the single-slit experiment) and the role in the UP of this kind of disturbance has not been studied by Brown & Redhead.

Now the interpretation of the uncertainty relation has been debated by many authors, but it has never been settled completely, and the debate actually resurfaced recently after Storey *et al*. [5] concluded that the interference pattern in a double-slit experiment is a consequence of Heisenberg's uncertainty relation and not of a more fundamental concept. The debate started earlier with a proposal from Scully *et al*. [6] where they theoretically showed that the disappearance of the interference pattern is due to entanglement and not to any "momentum kick" disturbance on the quantum system (see also Knight [7]). The debate continued by Englert *et al*. [8] and Storey *et al*. [9], and in 1998 the results of an experiment from Dürr *et al*. [10] were published, an experiment that seems to confirm what was theoretically predicted by Scully. Noteworthy is that the very recently performed experiments of Bertet *et al*. [11] also seems to confirm the predictions of Scully. A careful study of any classic interference experiment, like the double-slit (or single-slit as in Fig. 1), clearly reveals that upon passage through the slit(s) the quantum systems have indeed experienced some kind of a (transverse) momentum disturbance (or deviation) as the interference pattern exhibits a transverse momentum distribution along the screen. But, as Knight might put it, how exactly does the notion of this kind of disturbance get involved? As the definition of "disturbance" usually differs from author to author, also here one could ask: what exactly do the authors in the recent debate mean with "momentum" or "momentum kick" disturbance? Do they mean that the absolute value of the momentum, i.e., |**p**|, of the quantum system changed in their "which-way" experiment? Or do they mean that after passage through the slit(s) only the transverse momentum (distribution) of the quantum system has been altered or disturbed while tacitly ignoring any assumption on the new value of |**p**|. For instance, in the paper of Dürr the momentum disturbance they consider seems clearly to be of the latter type, i.e., a transverse momentum disturbance superimposed on the transverse momentum distribution. The central question that is dealt with in that recent debate is: Can their (transverse) momentum disturbance cause the interference pattern to vanish during a "which-way" experiment? The answer of Dürr is negative, i.e., the disturbance is too small and their conclusion is that the entanglement between the atom's momentum and its internal state destroys the interference. The question that I would like to put forward is an equally fundamental but a very different one, that is: to what extent does a transverse momentum disturbance (or deviation) play a role in the *build-up* or *creation* of the interference pattern during the time the quantum systems pass the slit(s) or right after, i.e., during the state preparation of the quantum systems.

In this paper I will try to answer the latter question by studying and defining the exact role of "disturbance" during the state preparation in the context of the single-slit experiment. So, looking at the disturbance problem from Ballentine's paper, I will pick up in the next two sections the thread where Ballentine actually left it, i.e., at the state preparation.

## 2. The single-slit experiment

A mathematical analysis of the single-slit experiment can, among other papers and books, be found in Hilgevoord [12], Uffink [13] and Jammer [14]. In this section I will focus on trying to see what actually happens on the level of individual systems (particle's). It is generally accepted that the diffraction curve in Fig. 1, which can be seen, for instance, on a (photographic) detector sheet (screen), is the result of the wave aspects of the individual systems. This means that when a single individual system passes trough the single-slit, the probability to detect the system somewhere on the detector is determined by the local amplitude of the diffraction curve. Once a first individual system has hit the detector a second system can be fired into the single-slit, after which it is then detected at some other position on the detector. This way a diffraction (interference) pattern gradually builds up as can be seen in the papers of Tonomura *et al*. [15] and Merli *et al*. [16] for the case of a double-slit experiment with electrons. Noteworthy to mention is that the experiments of Merli and Tonomura are biprism experiments. For a classical single- and double-slit experiment with electrons I would like to refer the reader to the very illustrative and groundbreaking work of Jönsson [17] published as early as 1961. Further in the paper I will refer mostly to the pictures in Fig. 5 of Tonomura's paper because those illustrate very clearly the statistical nature and the gradual build-up of the interference pattern. Also noteworthy to mention is that all diffraction and interference experiments referred to in this paper are wide-beam experiments. That is, in Fig. 1 the incident beam is at least as wide as the slit so that the corpuscular particle-part of the passing system(s) has an equal probability to appear anywhere within $\Delta x$.

Let us see what happens when only a few single systems pass through a vertically lined up single-slit and hit the detector. Note that the position of the impact of those few systems (electrons) on the detector can be determined very accurately (see Fig. 3 in Ballentine's paper and Fig. 5.a,b,c in Tonomura's paper). The systems detected on the detector will be at first sight scattered randomly to the left and the right. It is very clear without any doubt that the direction of motion of the few single systems has been changed or disturbed (there is now a transverse component, parallel along the screen) during some time between the time instance the systems entered the slit and the moment just before they impacted the detector, i.e., during the time of state preparation. This change or deviation in direction of the system's motion is what I consider and define as "disturbance" throughout this paper and which is essential for the observation of any diffraction or interference curve. This disturbance is quite different from the disturbance that for instance Dürr considered in his experiment. Furthermore, the single-slit is a massive and stiff device so that the net momentum transfer from the quantum system to the single-slit device upon the system's passage through the slit is negligible. This implies that we can assume that the absolute value of the system's momentum, i.e. |**p**|, remains practically unchanged during the state preparation.

As the number of systems detected on the detector increases from a few to a large number we notice that the disturbance in direction, which the individual systems have experienced, is not 100% random as the systems have a preference for certain directions and try to avoid others so as to build up a diffraction pattern with maximum and minimum density. As it is shown by Ballentine that the disturbance caused by the measurement itself cannot explain $\Delta x \Delta p \approx h$, it is obvious that the creation of the diffraction pattern and consequently the disturbance of the direction of the systems' motion must have taken place during the state preparation. Summarizing, we can say that during the state preparation of the systems there are two things at work here; (1) a disturbance in direction so that the systems get left and right scattered and (2) the in direction disturbed systems seem to be drawn or directed into certain directions which lead later on to the observed diffraction maximums on the detector. Therefore, the relation $\Delta x \Delta p \approx h$ has two aspects, i.e. first, the transverse momentum dispersion (curve (b) in Fig. 1) on the measured transverse momentum of an ensemble of systems and, second, the fact that $\Delta x$ and $\Delta p$ are complementary. We will first focus on the issue of $\Delta x$ and $\Delta p$ being complementary.

Let us have a look at the experimentally observed diffraction curve (a) in Fig. 1 and let us especially concentrate on the distance *W* between the two first-order minimums, which is proportional to $\Delta p$. Note that in Fraunhofer regime *W* can be approximated as

$$W \approx 2\lambda L / \Delta x \qquad (2)$$

where $\lambda$ is the wavelength of the systems, $\Delta x$ is the width of the slit and *L* is the distance between the single-slit and the screen. If we now decrease the width $\Delta x$ of the single-slit, then *W* will increase and vice versa, i.e., $\Delta x$ and *W* act complementary. As the position of the two first order minimums, and so the width *W* between them, is determined exclusively by the wave properties of the incident system, $\Delta x$ being complementary with *W* is therefore a pure wave phenomenon. Consequently, as $\Delta p$ is proportional to the distance *W*, we come to the conclusion that $\Delta x$ is complementary with $\Delta p$ is just as well a pure wave phenomenon. Let it be clear that the notion of $\Delta x$ and $\Delta p$ being complementary is not at all the result of any disturbance coming from the measurement itself on the screen nor is it the result of any disturbance (or deviation) in the system's direction of motion during and/or after passage through the slit.

However, with the aspect of the transverse momentum dispersion the situation is different and disturbance of the system's direction of motion does play a role. It is due to this disturbance that the systems get left and right scattered during the state preparation. That this disturbance is a prerequisite for observing a diffraction (interference) curve can be easily verified with the following though-experiment. Just image the situation where there isn't any disturbance in the system's direction of motion during the state preparation. This would imply that after passage through the slit the direction of the trajectories of all systems would be the same as the direction of the incident beam, i.e., along the symmetry line of the experimental setup. Such an experiment would result in the observation of a thin line as thick as the width of the slit and no diffraction pattern would be observed. So, as a conclusion we can say that the disturbance (or deviation) in transverse direction of the system's motion *during* state preparation is really *primary* and a *prerequisite* for introducing a transverse momentum spread (and a momentum dispersion curve, as curve (b) in Fig. 1) which on turn is essential for observing a diffraction curve.

There is, however, still the question of where exactly between the slit and the screen (and during the state preparation) curve (b) appears and where exactly the formation of the interference fringes (i.e. the clustering of the systems into certain directions) takes place? For the double-slit experiment, there is no doubt that the formation of the fringes cannot take place in the two slits itself nor can it take place right after the double-slit. The onset of the formation of the fringes must take place from some distance onwards behind the double-slit once the systems passed it. This can be deduced from Fig. 2 in Tonomura's paper. There is no reason that prohibits us from applying this finding to the single-slit case, that is: the formation of the fringes only happens once the systems have past the single-slit. Furthermore, curve (b) represents a transverse momentum spread that reflects the disturbance of the system's direction of motion. It is reasonable to assume that this disturbance is caused by some kind of an interaction between the system(s) and the edges of the slit. Therefore, the formation of curve (b) must happen during the time the systems are in the single-slit and/or maybe right after they passed the slit but certainly not at any time later. So, in this particular single-slit scheme that I envisage, there is first the formation of curve (b) which is effectively present right after the slit, and then later the superposition of the fringes on that curve so that curve (b) gradually evolves to curve (a) over some distance between the slit and the screen. Where exactly between the slit and the screen the formation of the fringes happens can be experimentally determined by varying the distance $L$ between the slit and the screen from Fraunhofer regime to a value in the order of $\Delta x$ or even less and see at what distance the interference fringes disappear. Note that from the technical point of view this experiment would require a high-resolution detector, or at least with the resolution that is a lot higher than $\Delta x$. The discussion in this paragraph should give the reader a clear picture about what the role is of the disturbance of the system's direction of motion in this single-slit scheme that I envisage.

Aside from the central issue of the role of disturbance in $\Delta x\, \Delta p \approx h$, an interesting question might be; what causes the disturbance in direction on the individual systems upon passage through the single-slit? Clearly, there seems to be an interaction somehow between the quantum system and the edges of the slit. The answer I would like propose is maybe controversial. First, it should be emphasized that all systems that pass the slit have a dual particle-wave character. Note that the notion of "wave" in this context means de Broglie wave. The still rudimentary particle-wave model that I envisage for the quantum system is one in which the system can be considered as a point-particle, which has a corpuscular (point-like and classical) character, joint by (or attached to) a wave that remains at all times in the immediate vicinity of the corpuscular particle-part and from which the amplitude or strength rolls off with the distance between the object with which there is an interaction (in this case the edges of the slit) and the particle-part of the quantum system. In other words, the further away the observer is from the particle-part, the weaker the observed wave is of the corresponding system. It is then the wave itself (and not the particle-part) that directly interacts with the edges of the slit during passage. Consequently, since the particle-part is attached to its corresponding wave, this direct interaction between wave and slit in turn causes the disturbance in the motion's direction of the particle-part of the system. As is shown in Fig. 5.a,b,c of Tonomura's paper, it is then this particle-part of the systems that one can see left and right scattered on the screen. The fact that it must be the wave and not the particle-part of the systems that directly interacts with the edges of the slit can be deduced from the fact that the incident beam of single systems can be taken as spatially homogeneously distributed across the slit during passage. That way the particle-part of the vast majority of the systems will pass the slit *between* the edges and not at or near the edges. From the interference experiments not only with photons but also with massive systems, such as electrons, neutrons (Zeilinger [18]), atoms and even molecules (Arndt *et al*. [19]), it is clear that this "disturbance in direction" must be caused by a classical mechanical force which pushes the systems to the left or right upon and/or (right) after passage through the slit.

Hence, the issue of the disturbance of the direction of the system's motion is confined to an interaction between the wave itself and the edges of the single-slit. Nevertheless, we come back to a similar question as the one that Knight came forward with, that is: how *exactly* does the wave interact with the edges of the slit? And, furthermore, how would the relation then be between that interaction and the direction of the system's motion after it passed the slit. If we can get a grip on that interaction, can we then also get a control on the disturbance and quantify it? Furthermore, it is noteworthy to mention that in this discussion assumptions have been made on two levels. First, in the rudimentary particle-wave model that I envisage and, second, in the single-slit scheme, i.e., I adhere the scheme in which first a dispersion curve is created on which then later the fringes are superimposed. Question: is that scheme really correct? Could it be that the momentum dispersion is not created in and/or right after the slit but further behind the slit simultaneously with the formation of the fringes? There is only one way to shed more light on the possible answers to all those questions in this paragraph, and that is by performing a novel experiment, which I will now propose in the following section.

### 3. Proposed experiment

The proposed experiment is given in Fig. 2. The essential issue in this experiment is that the (corpuscular) particle-part of all the quantum systems enters the slit in a very narrow range $b$ centered at one particular coordinate $x_b$. The fine beam of systems is composed of single system, i.e., a system enters the slit only when the preceding system has been detected. The systems in the beam should be mono-energetic, or at least as much as possible mono-energetic determined by the resolution of the experimental setup. The beam's thickness $b$ should be significantly smaller than the system's wavelength $\lambda$ and significantly smaller than the width $\Delta x$ of the slit. I would like to emphasize the fact that $b$ is determined by the particle-part of the systems and not by the wave that accompanies the corresponding particle-part. Just as in Fig. 1, in Fig. 2 only the particle-part of the systems is depicted and not the accompanying wave. So, in Fig. 1 and 2 the size of the black dots representing the particle-part of the systems is many orders smaller than the wavelength of the corresponding systems. Furthermore, the length $L$ between the slit and the screen is large enough so that we can consider the experiment in Fraunhofer regime.

Once the systems passed the slit at coordinate $x_b$ one might expect that on the screen a diffraction pattern with interference fringes (curve (a)) will build up with time. However, according to my insight there is a very good chance that curve (a) will not be the pattern that will be observed if the proposed experiment is implemented. I believe another pattern might appear, more specifically curve (c) in Fig. 2, that is, one narrow curve (line) at coordinate $x_p$ on the screen with a width $d$ significantly narrower than the width of an interference fringe, i.e., $d < \lambda L / \Delta x$. The qualitative explanation I give to curve (c) is the following. If an ensemble of identical systems all enter the slit at the same coordinate $x_b$, then upon passage through the slit all systems will experience exactly the same "disturbance in direction" because the relative position of the wave with respect to the edges of the slit is the same for all the passing quantum systems. That means that upon passage through the slit the systems are not randomly scattered but are all scattered in one particular and the same direction impacting the screen at coordinate $x_p$. According to my insight in the particle-wave model and in the single-slit experiment, the width $d$ of curve (c) will be a clear function of $L$, $\Delta x$, $b$ and $x_b$, i.e.,

$$d = f(L, \Delta x, b, x_b) \qquad (3)$$

where the position $x_p$ on the screen will be a function of $L$, $\Delta x$ and $x_b$, i.e.,

$$x_p = f(L, \Delta x, x_b). \qquad (4)$$

So if in Fig. 2 $x_b$ slightly increases then $d$ and $x_p$ should increase while if only $b$ slightly increases then only $d$ should increase. The role of $\Delta x$ in (3) and (4) is a bit more complicated, as will be discussed in the next paragraph, for the obvious reason that the position of the diffraction minimums are a function of $\Delta x$. For the specific case were $x_b = 0$, $x_p$ is then also zero because of geometric symmetry reasons of the wave relative to the edges of the slit. In the ideal and only theoretical case were $b = 0$, then according to my insight $d$ should also be zero, certainly at $x_b = 0$ (i.e. $x_p = 0$). This should give the reader an idea about what I believe could be expected if the proposed experiment would result in curve (c).

However, I expect that curve (c) will only appear as a narrow and single peak if $x_p$ coincides with the coordinates of the peak of one of the fringes of curve (a), i.e., coordinates … $f_{-2}$, $f_{-1}$, $f_0$, $f_1$, $f_2$, … in Fig. 2. If $x_p$ would coincide with the coordinate between two peaks in curve (a), then curve (c) will transform into a curve with two peaks as can be seen in Fig. 3. As in Fig. 3 the coordinate of the fine beam, i.e. $x_b$, sweeps from $x_b = 0$ to $0 < x_b < \Delta x/2$ the density curve will evolve from curve (d) over (e), (f), (g) to the curve (h). The fact that in Fig. 3 the double peaked curves have a density that is zero at positions $z_{-1}$ and $z_{-2}$ can be explained as follows. The interference minimums that can be seen in curve (a) are the result of the wave character of the individual quantum systems. So, whether we fire a single quantum into the slit as a stand alone system, or as part of a fine beam or as part of a wide beam as in Fig. 1, any quantum system will have a probability of near zero to hit the screen at coordinates …, $z_{-1}$, $z_{-2}$, $z_1$, $z_2$, … Therefore, in curve (e) and (f) the forbidden trajectory $x_b \rightarrow z_{-1}$ will split into the nearest allowed trajectories $x_b \rightarrow f_0$ and $x_b \rightarrow f_{-1}$ resulting in two density peaks.

If an implementation of this proposed experiment would result in the observation of curve (c) then not only would this experiment contribute to a better understanding of how interference is created and would it give us a deeper insight in the notion of "disturbance in direction" when a quantum system passes a slit; the experiment would also show that one can make predictions significantly more accurate than what Heisenberg suggests with his relation $\Delta x\, \Delta p \approx h$. That is, even before an experiment is performed one could from the values of $L$, $\Delta x$, $b$, $\lambda$ and $x_b$ in Fig. 2 predict the transverse momentum with an error $\Delta p$ so low that $\Delta x\, \Delta p$ would be significantly less than $h$: i.e. $\Delta x\, \Delta p \ll h$. The latter would also entail that Heisenberg's relation (and its corresponding uncertainty *principle*) is not fundamental in quantum physics but merely the result of the fact that quantum theory is an incomplete theory. On the level of the contemporary known quantum theory, $x_b$ would then be a hidden variable being made accessible thanks to the sub-wavelength diameter beam.

Question now is, how well can the proposed experiment be implemented? With neutrons as quantum systems that may turn out to be very difficult for the obvious reason that a neutron's trajectory is not so easy to control. So focusing a beam of low-energy neutrons into a beam with $b \ll \lambda$ ($\lambda$ = neutron's wavelength) could be a (too) difficult job. Atoms (ions), electrons and photons, on the other hand, should be easier to control and are therefore likely better candidates for using them as quantum systems in the experiment. Consider a single-slit electron experiment (Fig. 2) with the following dimensions: $\Delta x = 20\,\mu$m, $L = 1$ m, electron's wavelength $\lambda = 1\,n$m (10 Å) equivalent to 1.5 eV and $b > \Delta x$, then using expression (2) this results in $W = 100\,\mu$m. Suppose now that for the occasion of this paragraph we have an electron beam at our disposal with $b \ll \lambda$, let us say $b = 0.1\,n$m (1 Å) or even less, then curve (c) should appear on the screen with $d$ significantly smaller than $W$ for particular values of $x_b$, if at least the resolution of the photo sheet or the screen is high enough. If this resolution would be too low then one can always expand curve (c) by introducing an electron-optical enlargement between the slit and the detection system just as Jönsson and Tonomura did in their experiments. So the width $W$ or $d$ should not pose any technical difficulties from the viewpoint of the screen's resolution. As we shall discuss in the next section, the technical challenge, however, does not lay in the single-slit experiment itself but lays in generating a beam with sub-wavelength diameter, i.e., with $b \ll \lambda$.

### 4. Sub-wavelength beam diameter

One may raise questions about the feasibility of such a beam with sub-wavelength diameter and wonder whether or not that is even possible in principle? And even so, is the basic technology available to develop such a sub-wavelength diameter beam? Well, it is my believe that both questions can be answered affirmative and the discussion in this section will further elucidate this issue of feasibility.

Focusing a beam so that the particle-part of all quantum systems hits the target on an area from which the diameter is significantly smaller than the wavelength is in principle possible. Let us take the following thought experiment. A single photon particle of wavelength $\lambda = 600\,n$m (visible range), corresponding with an energy $E \approx 2$ eV, impinges a target plate composed of atoms from which the ionization energy is $\approx 1.9$ eV.

Note that just as a bound electron from an excited atom can emit a photon after de-excitation, the same electron can absorb a photon. So, in the best case the impinging photon will release no more than one electron upon its absorption and therefore ionize only one atom. The fact that only one electron is released, which was before confined in a space from which the diameter is at most 1 Å (0.1 $n$m) while the photon's $\lambda = 600$ $n$m, is a clear indication that all the photon's energy must have been confined in a very small space which is at least significantly smaller than the photon's wavelength, and that is in the photon's particle-part. Once the ionized ion has recombined, and so back neutral, there is no objection to fire a second 2 eV photon towards the recombined atom so that its particle-part interacts with the same atom and ionizes it again. Some time later it recombines again and we can repeat the procedure many more times. This, in fact, corresponds with a 2 eV photon beam from which the photon's particle-part falls within a beam diameter of 1 Å or less. So, in principle there is no problem at all with firing a 1 Å 2 eV photon beam towards the same atom.

An issue to pay attention to maybe is how to define a "beam" diameter. In Fig. 2 of this paper the beam is defined by the particle-part of the quantum systems. In some experimental setups, however, the beam is defined in a different way. For instance, in the field of Scanning Electron Beam Microscopy, the beam diameter is sometimes defined as a certain percentage of the maximum of the impacting current at the target (see Weyland [20] Chapter 3, and Refs. 21, 22 and 23). It is not unimaginable that in an optics experiment the quantum systems in the beam are considered as spheres from which the diameter is one, two, or even more wavelengths. So, in such an experiment it is then quite trivial that a beam made up of the latter quantum systems can have a diameter of no less than respectively one, two or more wavelengths depending on the minimum size considered for the quantum systems. In our single-slit experiment the minimum size of the quantum systems is the size of their particle-part (so any wave excluded) so that the beam diameter can be significantly less than the wavelength of the quantum system in the beam.

Now that it is clear that a beam diameter of less than the wavelength is in principle possible, the question remains if also the technology is available to develop such a thin beam? Well, the basic technology to develop such a beam using different quantum systems such as photons, atoms and electrons, is definitely available. Let us consider the situation for photons. In the field of optical telecommunications single-mode optical fibers are used to transport the optical signals. Those fibers allow a single frequency (channel) single-mode optical signal to travel over distances of several hundreds of kilometers with almost no dispersion. According to Ishak [24], theoretically a single-mode optical fiber could carry 25 terabits per second. In practice the (opto)electronics at the fiber-ends is the limiting factor; currents state of the art is in the order of 40 Gbit/s per single frequency (per channel) single-mode optical signal [25]. More information about how single-mode optical fibers transport optical signals and about the issue of (wave-guide) dispersion can be found in Refs. 26 and 27. The fact that the dispersion of a single frequency single-mode pulse is so small is due to the fact that the particle-part of the photons moves along the central axis of the optical fiber within a range in the order of less than a fraction of its wavelength. Hence, certainly in a straight stretched fiber, the particle-part of the photons make up a beam from which the diameter is only a fraction of the wavelength. If that would not be the case and suppose the photons' particle-part would move along different parallel trajectories further away from the central axis in the fiber, then even the slightest bending of the fiber would cause a significant extra dispersion, greatly degrading the performance of the fiber, and that is clearly not the case in single-mode fibers.

Question now is, would it be possible to extract this photon beam from the fiber and pass is trough the single-slit of our experiment? Of course, as we cannot pass the fiber itself through the single-slit, we have to break off the fiber so that the fiber ends with a tip. A possibility is then to position the fiber's end (fiber tip) right in front of the single-slit. Note that one should pay attention to the fact that once the photon beam has left the fiber, that the beam doesn't diverge but remains parallel so that it's diameter remains only a fraction of the beam's wavelength. The latter would at least require a particular shaped fiber tip and maybe even extra focusing optics.

One might wonder if a beam with a sub-wavelength diameter would allow us to probe structures with sub-wavelength resolution? Well, the answer is definitely yes and such a device already exists for more than a decade and is called the Scanning Near-field Optical Microscope (SNOM) [28-31]. Laser light in the visible range ($\lambda = 600$ $n$m to $400$ $n$m) is fired into a single-mode optical fiber and then focused onto a spot with a diameter in the order of $50$ $n$m, or even less [29], using a coated fiber tip. The resolution varies from $\lambda/10$ to $\lambda/15$ and even higher depending on the fiber tip technology. There is no doubt that the basic technology to focus photons on a spot with a sub-wavelength diameter is available since more than decade. The only problem with a SNOM device is that the direction of part of the exiting photons at the fiber tip is not parallel with the central axis of symmetry of the fiber tip, i.e., the photon beam diverges. So, although the basic technology is commercially available, there is still some technological development ahead in order to ensure a beam of parallel moving photons once they left the fiber tip.

Now, a photon beam might look obvious but photons are actually an extreme case in that respect that they do not have a rest mass. Let us go to the other extreme and discuss the situation of a beam of atoms, more specifically of atom ions. Laser cooling and trapping of a single ion can be performed in several different ways but I would like to focus on a combination of a linear Paul trap with laser cooling [32-36]. State of art trapping and cooling manages to confine a single ion within a space with dimensions smaller than $10$ $n$m (see Ref. 37, p. 263 and Ref. 38, p. 8), i.e., the width of the single ion's wave function is less than $10$ $n$m. This is already good but not really good enough for our single-slit experiment. The way to confine the single ion even more is by introducing an optical cavity, thereby introducing cavity Quantum Electro Dynamics (QED). Recent work from Gulde *et al.* [39] (Section 2.2.2) reports about a localization down to $7$ $n$m for the ions in a one-dimensional lattice and Eschner *et al.* [40] (p. 367) reports about a confinement of 6 $n$m, and that in axial direction for both publications. Thanks to the ion-cavity mode coupling the ion(s) can be exactly positioned with $n$m accuracy. This $n$m positioning accuracy (radial, however) is needed to generate a sub-wavelength beam diameter for our single-slit experiment. The required basic technology is already available in a number of laboratories around the world.

Now, how exactly would our proposed experimental setup with trapped ions look like? Well, we start with a vertically aligned linear Paul trap, i.e., the central axis of the trap is vertical (and parallel along the direction of the earth's gravitational field lines) as in Fig. 2 of Ref. 35. We call this vertical (central symmetry) axis the $z$-axis and in radial (horizontal) direction we then have the $x$- and $y$-axis. Let us go back to Fig. 7 of Ref. 39 or Fig. 5 of Ref. 40. Both mirrors of the optical cavity are spherical and the focal point of the cavity lies on the central axis of the Paul trap. Note also that in the schematic experimental setup in their Figures the central axis of the optical cavity makes an angle with the central axis of the Paul trap. In our proposed experiment this angle must be 90 degrees so that the extra strong confinement is established in radial direction, more specifically in the $x$-direction. Furthermore, in our suggested experiment the mirrors need to be cylindrical (in stead of spherical) with a height of, let us say, 2 $c$m. What was a focal point before with spherical mirrors becomes now with cylindrical mirrors a focal line parallel to the two mirror surfaces. This focal line must also coincide with the central axis of the Paul trap, i.e., the $z$-axis. This way a strong confinement in the $x$-direction, in the order of a few $n$m, can be established over a distance of 2 $c$m along the $z$-axis. The single ion will then be trapped and cooled exactly in the middle of the cavity, with an accuracy of $n$m in the $x$-direction and in the order of less than a micrometer ($\mu$m) along the $z$-direction. Based on the literature, I expect that a high performing classical laser cooling in the $z$-direction is sufficient to obtain a single ion positioning with $\mu$m accuracy and that an optical cavity (cooling) in the $z$-direction in not needed.

The idea is then the following. Mount right underneath the optical cavity a horizontally positioned single-slit plate from which the slit width is, let us say, 200 to 400 $n$m and is aligned along the $y$-axis, i.e. perpendicular to the direction of the strong optical cavity confinement which is in the $x$-direction. The (central) $z$-axis of the trap must pass through the single-slit. The next step is then to trap and cool a single $Ca^+$ ion in the middle of the optical cavity so that the difference in height ($h$) between the position of the cooled $Ca^+$ ion and the single-slit plate is $h = 1$ $c$m. Once the $Ca^+$ ion is maximally cooled, radial as well as along the $z$-axis, one can then switch off only the vertical cooling, while maintaining the radial cooling and maintaining the strong (optical cavity driven) confinement in the $x$-direction.

The result is that the $Ca^+$ ion drops in free fall over a distance of 1 $c$m and thereby gains a kinetic energy $E_k$, which corresponds with a certain wavelength for the ion by the time it reaches the single-slit. Note that the idea of letting an atom or ion fall freely is also used in Cesium Fountain Atomic Clocks [41]. Once the $Ca^+$ ion reaches the slit, and that is after $t = \sqrt{(2h/g)} \approx 45$ $m$s with $h = 1$ $c$m and $g \approx 9.81$ m/s$^2$, then ALL cooling must be switched off so that the ion can freely interfere with the single-slit while passing it. Some distance underneath the single-slit plate we could mount a horizontal substrate on which the ion can then drop onto and remain stuck to the substrate without diffusing or wondering around on the substrate. Once the first $Ca^+$ ion has reached the substrate then the whole procedure of loading the trap with a second $Ca^+$ ion, further trapping and cooling it repeats itself resulting in a second $Ca^+$ ion that reaches the substrate underneath the single-slit. If the procedure is repeated many times, an atom (interference) pattern will emerge on the substrate. Later on we can then analyze the substrate surface and observe the atom (interference) pattern by using, for instance, a Scanning Electron Beam Microscope. Noteworthy to mention is that the atom pattern on the substrate will be deformed due to the gravitational acceleration the $Ca^+$ ions keep on experiencing once they passed the single-slit. This deformation can be easily calculated.

What now follows are some key figures. The atom weight of Ca is $\approx 40$ so that the mass $m$ can be calculated as $m \approx 6.7 \times 10^{-26}$ kg. The kinetic energy $E_k$ of the $Ca^+$ ion at the single-slit equals the potential energy $E_p$ of the $Ca^+$ ion for a height ($h$) of 1 cm. Hence, with $g \approx 9.81$ m/s$^2$, $E_k = E_p = mgh \approx 41$ neV. From this $E_k$, and with $E_k <<< E_o$, one can calculate the ion's wavelength $\lambda$ using the approximate relation $\lambda \approx hc/(\sqrt{(2E_k E_o)})$, where $h$ is Planck's constant, $c$ is the speed of light and $E_o$ is the rest energy of the $Ca^+$ ion being $\approx 38$ GeV. This results in $\lambda \approx 21$ nm, which is significantly larger than the $Ca^+$ ion's (along the $x$-axis) strongly confined beam width of only a few $n$m. Replacing the $Ca^+$ beam by a $Be^+$ beam would give us a $\lambda \approx 105$ nm. Furthermore, in the setup that Nägerl uses (Ref. 34, p. 4; Ref. 35, p. 626), the Doppler limit is around $T = 0.5$ mK with quantum numbers $\langle n \rangle_r = 7$, $\langle n \rangle_z = 77$ and with trap frequencies of $\nu_r = 1.39$ MHz and $\nu_z = 134$ kHz. If one would apply resolved-sideband Raman cooling [42], then the $Ca^+$ ion can be cooled to the zero-point energy, given as $E = h\nu/2$. With aforementioned figures for the frequency this results in a zero-point radial energy $E_r \approx 2.9$ neV and a zero-point axial (vertical) energy $E_z \approx 0.3$ neV respectively corresponding with $\lambda_r \approx 83$ nm and $\lambda_z \approx 245$ nm.

Question is: are these energies low enough so to generate an $n$m-wide mono-energetic ion beam which is not too divergent. In other words, are the trajectories of the free falling ions parallel enough? Well, the vertical energy $E_z \approx 0.3$ neV is maybe low enough not to deteriorate the mono-energetic character of the ion beam, but the critical quantity here is the radial energy, $E_r$, (especially along the $x$-direction) and a value of $\approx 2.9$ neV ($\approx 4.6 \times 10^{-28}$ J) is far too high. This can be easily verified. Let us call the radial energy along the $x$-direction $E_x$, and its corresponding momentum $p_x$. With a single-slit width of 200 nm $p_x$ of the incident ion beam should remain significantly smaller than $h/\Delta x$, i.e., $p_x << h/\Delta x \approx 3.3 \times 10^{-27}$ N.s, in order not to destroy the expected patterns in Fig. 3. Let us put forward a zero-point $p_x$ two orders of magnitude smaller than that limit, say $p_x = 10^{-29}$ N.s; with $m$ the mass of the $Ca^+$ ion, this then corresponds with the critical limit $E_x = p^2/2m \approx 7.7 \times 10^{-34}$ J which is 6 orders of magnitude smaller than the 2.9 neV we have with the trapping conditions in the Paul trap from previous paragraph.

Note that in the case of Ca *atoms* (in stead of Ca$^+$) the $E_x$ of the incident atoms in the single-slit would be determined not only by their secular ground state motion in the trap but also by their temperature, $T$, they have when several atoms would be at the same time in the trap (case of dilute Bose-Einstein condensate (BEC)). With $E = k_B T/2$ one can calculate that an $E \approx 7.7 \times 10^{-34}$ J corresponds with a $T \approx 100$ pK. Although next reference is in the field of BEC using a gravito-magnetic trap and not in the field of single ion trapping, Ketterle's group [43] reported effective temperatures in the range of 30 pK for a BEC with 2500 sodium ($^{23}$Na) atoms. The other problem is to reduce the trap frequency. The zero-point energy must be in the range (or lower) of $E \approx 7.7 \times 10^{-34}$ J which corresponds (using the relation $E = h\nu/2$ ) with a trap frequency of $\nu \approx 3$ Hz. The trap frequency was in the case of [43] as low as 1 Hz. So, the basic technology for creating temperatures in the pK range and trap frequencies in the range of 1 Hz (or lower) is available. Does this mean that in principle we could drop with nm accuracy mono-energetic ions/atoms one after another at always the same x-coordinate in a single-slit? Well, not quite (yet). From any quantum physics text book discussing the linear harmonic potential oscillator, the width (FWHM) of the zero-point eigenfunction $\psi_0(x)$ at $\nu = 1$ Hz can be calculated as $2(\sqrt{(-2\ln(1/2)/\alpha^2)}) \approx 40\,\mu$m with $\alpha = m(2\pi)^2\nu/h$ and $m$ the mass of the Ca atom. Hence, at $z \approx 0$ the probability for finding the atom is spread along the x-direction over a range of at least 40 $\mu$m and that is a lot larger than the 200 nm single-slit width. Does this mean that atom/ion trapping is not suitable to create a sub-wavelength beam diameter? Well, that is still to be seen.

In any experimental setup – Paul or gravito-magnetic trap – the technological challenge lays in knowing where the ion/atom is located along the x-direction with nm accuracy before dropping it into the single-slit. This challenge or problem has much to do with the procedure that is followed: i.e. first cooling to zero-level at $\nu \approx 1$ Hz and then later trying to find out where the ion/atom is located by some measurement. In fact, there is also another method for creating a sub-wavelength diameter beam without having exact knowledge of where the ion/atom is in the trap: and that is by using an ion/atom-optical lens. Take now the situation where the ground state eigenfunction $\psi_0(x)$ stretches over a range of 100 $\mu$m (at $\nu \approx 1$ Hz or lower). After the vertical trapping is switched off, the ion(s) or atom(s) drop(s) down in free fall and could be focused onto an area of just a few nm wide (in the single-slit) using an ion/atom-optical lens [44] in a coaxial arrangement (see also Ref. 45 for more recent references on atom beam focusing). Once the ion(s) or atom(s) has (have) reached the slit, then the atom lens is also switched off so that the particle can interact/diffract freely with the slit. After the vertical trapping is switched off, it takes the ion/atom(s) a time of 45 ms to reach the slit. During that time the ion/atom(s) need to be shifted over a distance of in the worst case $\pm 50\,\mu$m from the outer left and right ends of the trap (or better $\psi_0(x)$) to the center of the trap. This leads to an approximate maximum velocity along the x-direction (ignoring the $\pm$ sign) of $v_x \approx 50\,\mu$m/45 ms $\approx 10^{-3}$ m/s and a corresponding extra transverse momentum $p_x \approx mv$ $\approx 0.7 \times 10^{-28}$ N.s which is lower than the upper limit of $\approx 3.3 \times 10^{-27}$ N.s found earlier in order not to destroy the expected patterns in Fig. 3. Nonetheless, the extra value of $\approx 10^{-28}$ N.s may not be low enough. If so, then I propose the following solution.

Consider a purely theoretical exercise with, for instance, a gravito-magnetic trap [43] with a limited number of Ca atoms in it (dilute BEC) at $v \approx 1$ Hz and with a zero-point $p_x$ for the atoms in the trap far below the upper limit of $\approx 3.3 \times 10^{-27}$ N.s. The atoms will be located in the center of the trap over a distance of, let us say, $100 \mu$m along the x-direction. The next step is then to fire in the y-direction a well-focused and short laser pulse with appropriate wavelength into the condensate so to knock-out the atoms out of the trap in the y-direction, except for those atoms that are within a few $\mu$m of the center in the trap so that the width of the condensate is reduced from $100 \mu$m to $2 \mu$m. The consequence for the quantum theoretical interpretation is that the eigenfunction $\psi_0(x)$ collapses instantaneous and (unfortunately only) temporarily to a Dirac pulse of $2 \mu$m wide, practically without changing the zero-point $p_x$ of the remaining trapped atoms. If, however, the collapse would be not temporarily but definite ($p_x \approx 0$)) then we could use an atom-optical lens to focus this remaining $2 \mu$m wide condensate to an area of a few $n$m wide in the single-slit. The maximum atom displacement would be only $1 \mu$m during the fall time of $45 m$s corresponding to an additional maximum $v_x \approx 22 \mu$m/s and $p_x \approx 1.4 \times 10^{-30}$ N.s which is far below the upper limit of $\approx 3.3 \times 10^{-27}$ N.s. Note that in the experiment of Ref. 43 not Ca but Na was used. In fact, some first principle calculations reveal that it might be easier to implement the experiment using Na instead of Ca. With a trap frequency $v \approx 1$ Hz and sodium mass $m(Na) \approx 3.8 \times 10^{-26}$ kg the maximum zero-point $p_x$ can be calculated as $\approx \sqrt{(2mE)} \approx 5 \times 10^{-30}$ N.s. In order to make the Na atoms acquire an $E_k \approx 41 n$eV by the time they reach the slit, they need to fall over a height $h \approx 2 c$m and that takes a time of $\approx 90 m$s. $\psi_0(x)$ will be spread over a distance of $\approx 200 \mu$m but this is of no relevance as only the $2 \mu$m wide region in the center of the trap is kept in the trap after use of the laser knockout procedure. With $h = 2 c$m, the atom-optical lens would then introduce a maximum $v_x$ of $\approx 12 \mu$m/s and so a $p_x$ of $\approx 4 \times 10^{-31}$ N.s which is lower than in the case of Ca atoms. Unfortunately the zero-point $p_x$ of the condensate is not $\approx 0$ and so the main problem still to be solved is the zero-point lateral drift of the atoms. Note that a zero-point $p_x \approx 5 \times 10^{-30}$ N.s corresponds with a maximum $v_x$ of $\approx 130 \mu$m/s so that by the time the sodium atoms reach the slit ($90 m$s later), the atoms may have shifted along the x-direction over a distance of maximum $12 \mu$m, far away from the slit. The solution that I see for this problem is a combination of measures: i.e. lower trap frequency, smaller height $h$ and especially sustaining the laser knockout procedure during a longer time. Let us take $h = 1 c$m for the case of Na atoms so that the fall time is $45 m$s. This corresponds then with a $E_k \approx 23 n$eV and a $\lambda \approx 37 n$m by the time they reach the slit. If we impose a lateral drift of just a few $n$m during the $45 m$s fall time then this corresponds with $v_x \approx 44 n$m/s. Next question is how long must the laser knockout then last in order to select those atoms with a maximum $v_x \approx 44 n$m/s? Let us take a width of $2 \mu$m in the center of the condensate that is not knocked out by the laser; all atoms which end up outside this $2 \mu$m region are then being removed. If the atom removal is sustained for a time of $2 \mu$m / $44 n$m/s $\approx 45$ seconds, then the remaining atoms in the trap will not only be known to be in the $2 \mu$m region but will also have a maximum $v_x \approx 44 n$m/s. After $45 s$ the (vertical) trapping (and cooling) can be switched off and the remaining atoms are then atom-optically focused into the single-slit. This laser removal procedure, which is sustained for $45 s$, entails not only a collapse of $\psi_0(x)$ – to a Dirac pulse of $2 \mu$m wide – but also a zero-point momentum ($p_x$) collapse. In conclusion, the explained procedure in combination with trap frequencies even lower than 1 Hz has in my view the potential for generating an atom beam with sub-wavelength diameter.

Another very promising technology for implementing an atom beam with sub-wavelength diameter is the Scanning Tunneling Microscope (STM). Already in 1991 Eigler *et al*. [46] succeeded in picking up single Xenon (Xe) and Platinum (Pt) atoms from a substrate using a STM tip and then deposit them back with Ångström accuracy at another location on the substrate by applying a voltage pulse to the STM tip so to eject the atom off the tip. For a more recent overview see also the work of Meyer *et al*. [47]. This STM single atom ejection could be fine tuned for the purpose of generating the necessary sub-wavelength diameter atom beam. Before doing this it is noteworthy to mention that the work done by Eigler and other research teams later required a current between the substrate and the STM tip in order to move the target atom (Xe, Pt) from the substrate to the tip and back. Hence, the substrate in their cases was electrically conducting. In the case of this single-slit experiment the slit itself is empty space and not conducting at all. Nevertheless, an atom at the tip of a STM can be ejected into empty space or onto an insulating substrate in a field-evaporative way: i.e., the atom at the tip can be evaporated or ejected away from the tip by use of a high amplitude voltage pulse. In field-evaporative mode the amplitude of the voltage pulse runs in the order of Volts while with a conducting substrate this voltage is a lot lower. As a good starting point, relatively recent research and interesting references in field-evaporation can be found in the work of Park *et al*. [48] and Tsong *et al*. [49].

The STM experiment I would like to propose is the following. A single Ca atom (or any other type of atom for that matter) could be picked up by the STM tip and positioned with *n*m accuracy above the 200 *n*m wide single-slit at a particular *x*-coordinate. The *z*-distance (height) between the STM tip and single-slit plate must be in the order of one or just a few $\mu$m. Then the atom is ejected downwards, so vertically and parallel along the *z*-direction, and passes the slit at a particular *x*-coordinate. The energy with which the atoms are ejected from the STM tip is roughly determined by the duration and the amplitude of the applied voltage pulse. Therefore, the pulse amplitude and duration can be adjusted so that this energy with which the atom is ejected from the tip should be approximately 41 *n*eV as calculated earlier. The same procedure repeats itself and the next atom that is picked up is then also ejected with about the same energy ($\approx$41 *n*eV) into the single-slit at the same *x*-coordinate as the previous atom(s). Underneath the single-slit there is the substrate onto which the atoms can fall and remain stuck to form the in Fig. 3 expected density profile. This is basically the proposed experiment. One issue to pay attention to is the fact that $p_x$ (and maybe also $p_y$) must be significantly smaller than $\approx 3.3 \times 10^{-27}$ N.s, a necessary condition for observing the expected curves in Fig. 3, and this may require extra research in the design of the STM tip. Nevertheless, in the field of STM the basic technology is definitely available to try to perform the proposed experiment. In fact, according to my view STM technology is probably the easiest technology to implement the single-slit experiment, although the experiment would need to be implemented in ultra high vacuum (UHV) and at a very low temperature ($<4$ K).

Up until now two types of experiments were studied: i.e. with photons who's rest mass is zero and with $Ca^+$ ions en Ca atoms with a (large) rest mass. However, one could also take quantum systems with a rest mass in between the two extremes, such as electrons. Let us have a quick look at Weyland's Ph.D. Thesis [20], Chapter 3, Fig. 3.7. In one of the sub-pictures an electron beam diameter of 0.28 nm at 3.99 keV is shown. At that energy the beam diameter is only 14 times larger than the electron's wavelength $\lambda = 0.02$ nm. Noteworthy to mention is that this beam is accomplished using the CM300-FEG-(S)TEM scanning e-beam device from Philips which is an ordinary commercially available technology and not even state of the art anymore. With exactly the same (S)TEM the beam diameter could even be significantly lower (with the same energy) if it wasn't from the additional magnetic coils and other things in the (S)TEM which is necessary for scanning the e-beam over the sample under investigation. Furthermore, there is also a technique to drastically decelerate the electron speed of the beam, called retarding field optics. Horden *et al*. [50] worked out an improved form of this technique so to avoid deformation of the beam during retardation. The possibility that I see is to fine tune and further develop this retarding technique so to retard, for instance, the above 0.28 nm-3.99 keV beam to an energy of, let us say, 0.1 eV which would correspond with $\lambda = 3.9$ nm. If then the beam diameter of 0.28 nm would remain unaltered, this would result in a sub-wavelength beam diameter for electrons.

In summary, I would like to emphasize that, although some further technological development needs to be done, the basic technology needed to develop an ion/atom, electron or photon beam with sub-wavelength diameter is definitely available.

Before coming to the conclusion in the next section, there is still one issue I would like to address. One might be tempted to think that just as the slit that is used to prepare a diffraction or interference pattern will modify the incoming beam according to the Uncertainty Principle, the device which prepares the beam must also modify the beam under preparation according to the Uncertainty Principle so that it would be impossible to have a sub-wavelength diameter beam at the slit. This line of thought is clearly wrong and for the following reason. Single-slit and double-slit experiments are typically de Broglie wave diffraction-interference experiments; i.e., there is no electrical nor any magnetic interaction between the slit and the quantum systems which pass the slit (see e.g. the case of neutrons in Zeilinger's paper). Even when the particle-part of the quantum systems is charged as in the case of ion's or electrons, the resulting interference pattern on the screen is a pure de Broglie wave interference pattern which clearly indicates that there is no electric nor any magnetic interaction between the charged particle and the slit plate (see also original paper of Jönsson [17] for technical details). Furthermore, my proposed single-slit experiment is intended to produce only a de Broglie wave diffraction, i.e., there is no electric nor magnetic interaction what so ever between the beam and the slit plate. Question now is how to generate and focus a beam (with sub-wavelength diameter) without interacting with the de Broglie wave of the quantum systems in the beam? Well, one possibility to accomplish this is with the aid of the quantum systems' charge or magnetic momentum (in case of ions, atoms and electrons) in combination with the use of external electric or magnetic fields because those fields interact only with the charge or magnetic momentum and not with the de Broglie wave of the quantum systems. In summary for this paragraph I would like to mention that in my view the de Broglie wave related Uncertainty Principle does not necessarily play a role, depending on the technical implementation of the beam preparer, in the under limit of the beam diameter during the beam's preparation.

## 5. Conclusions

There are a few important aspects about the relation $\Delta x\, \Delta p \approx h$ that I would like to summarize. First, the fact that $\Delta x$ and $\Delta p$ are complementary during the state preparation is due solely to the system's wave aspect and not to any disturbance during the measurement itself nor to any disturbance in direction (or in transverse momentum) during the state preparation. Second, the disturbance on the direction of the system's motion during the state preparation is a *prerequisite* for introducing the notion of a transverse momentum distribution and thus for observing a diffraction curve. This paper then proposes a single-slit experiment that, if my predicted outcome in Fig. 3 is correct, should give a deeper insight in the wave-particle duality and in the true interpretation of Heisenberg's uncertainty relation and its connection with the notion of disturbance. In the last section a feasibility study is undertaken for the development of a beam with sub-wavelength diameter. From this study it is clear that the basic technology needed for an implementation of the proposed experiment is available.

___________________________________________________________________________

## References

[1]  W. Heisenberg, Zeitschrift für Physik **43**, 172 (1927).

[2]  L.E. Ballentine, Reviews of Modern Physics **42**, 358 (1970).

[3]  C.A. Hooker, Australasian Journal of Philosophy **49**, 262 (1971).

[4]  H.R. Brown and M.L.G. Redhead, Foundations of Physics **11**, Nos.1/2, 1 (1981).

[5]  E.P. Storey, S.M. Tan, M.J. Collett, and D.F. Walls, Nature **367**, 626 (1994).

[6]  M.O. Scully, B.-G. Englert, and H. Walther, Nature **351**, 111 (1991).

[7]  P. Knight, Nature **395**, 12 (1998).

[8]  B.-G. Englert, M.O. Scully, and H. Walther, Nature **375**, 367 (1995).

[9]  E.P. Storey, S.M. Tan, M.J. Collett, and D.F. Walls, Nature **375**, 368 (1995).

[10] S. Dürr, T. Nonn, and G. Rempe, Nature **395**, 33 (1998).

[11] P. Bertet, S. Osnaghi, A. Rauschenbeutel, G. Nogues, A. Auffeves, M. Brune, J.M. Raimond, and S. Haroche, Nature **411**, 166 (2001).

[12] J. Hilgevoord and J.B.M. Uffink, European Journal of Physics **6**, 165 (1985).

[13] J.B.M. Uffink and J. Hilgevoord, Foundations of Physics **15**, 925 (1985).

[14] M. Jammer, The Philosophy of Quantum Mechanics: The Interpretations of Quantum Mechanics in Historical Perspective (A Wiley-Interscience Publication, 1974).

[15] A. Tonomura, N. Osakabe, T. Matsuda, T. Kawasaki, and H. Ezawa, American Journal of Physics **57**, 117 (1989).

[16] P.G. Merli, G.F. Missiroli, and G. Pozzi, American Journal of Physics **44**, 306 (1976).

[17] C. Jönsson, Zeitschrift für Physik **161**, 454 (1961), Translation in the American Journal of Physics **42**, 4 (1974).

[18] A. Zeilinger, Reviews of Modern Physics **60**, 1067 (1988).

[19] M. Arndt, O. Nairz, J. Vos-Andreae, C. Keller, G. Van Der Zouw, and A. Zeilinger, Nature **401**, 680 (1999).

[20] M. Weyland, *Two and Tree Dimensional Nanoscale Analysis: New Techniques and Applicatons*, (Ph.D. Thesis, Girton College, Cambridge, UK, Dec. 2001). http://www-hrem.msm.cam.ac.uk/~mw259/Thesis/Thesis.html

[21] J. Goldstein and M. Staniforth, *Scanning Electron Microscopy and X-Ray Microanalysis* (Kluwer Academic Publishers; 3rd edition, February 2003).

[22] S.L. Flegler, J.W. Heckman, and K.L. Klomparens, *Scanning and Transmission Electron Microscopy* (Oxford University Press, Oxford, UK, June 1997).

[23] J.R. Michael and D.B. Williams, Journal of Microscopy-Oxford **147**, 289 (1987).

[24] http://www.kth.se/fakulteter/TFY/kmf/Waguih.html
http://en.wikipedia.org/wiki/Optical_fiber

[25] R. Ramaswami, *Optical Fiber Communication: From Transmission to Networking*, IEEE Communications Magazine, 50th Anniversary Commemorative Issue (May 2002).
http://www.comsoc.org/livepubs/ci1/public/anniv/pdfs/rama.pdf

[26] J.A. Buck, *Fundamentals of Optical Fibers* (Wiley-Interscience, March 1995).

[27] http://www.fiber-optics.info/glossary-a.htm

[28] http://www.olympusmicro.com/primer/techniques/nearfield/nearfieldintro.html
http://monet.physik.unibas.ch/snom/

[29] E. Betzig, J. Trautman, and T. Harris, Science **251**, 1468 (1991).

[30] G. Tarrach, M.A. Bopp, D. Zeisel, and A.J. Meixner, Review of Scientific Instruments **66**, 3569 (1995); http://rsi.aip.org/rsi/

[31] B. Hecht, *Forbidden light Scanning Near-Field Optical Microcopy*, (Ph.D. Thesis, Basel, Switserland, 1996); http://www.nano-optics.ch/docs/diss.pdf

[32] J. Eschner, G. Morigi, F. Schmidt-Kaler and R. Blatt, J. Opt. Soc. Am. B **20**, 1003 (2003); http://heart-c704.uibk.ac.at/publications/papers/josab03_eschner.pdf

[33] H.C. Nägerl, *Ion Strings for Quantum Computation*, (Ph.D. Thesis, Innsbruck, Austria, September 1998).
http://heart-c704.uibk.ac.at/publications/dissertation/hcn_diss.pdf

[34] H.C. Nägerl, C. Roos, D. Leibfried, H. Rohde, G. Thalhammer, J. Eschner, F. Schmidt-Kaler, and R. Blatt, Physical Review A **61**, 023405-1 (2000).
http://heart-c704.uibk.ac.at/publications/papers/pra00_naegerl.pdf

[35] H.C. Nägerl, C. Roos, H. Rohde, D. Leibfried, J. Eschner, F. Schmidt-Kaler, and R. Blatt, Fortschritte der Physik **48**, 623 (2000).
http://heart-c704.uibk.ac.at/publications/papers/fortschritte00_naegerl.pdf

[36] see Refs 32, 33, 34, and 35

[37] D.J. Wineland, C. Monroe, W.M. Itano, D. Leibfried, B.E. King, and D.M. Meekhof, J. Res. Natl. Inst. Stand. Technol. **103**, 259 (1998).
http://nvl.nist.gov/pub/nistpubs/jres/103/3/j33win.pdf

[38] M. Sasura and V. Buzek, Journal of modern optics **49**, 1593 (2002).
http://www.quniverse.sk/buzek/mypapers/02jmo1593.pdf

[39] S. Gulde, H. Häffner, M. Riebe, G. Lancaster, A. Mundt, A. Kreuter, C. Russo, C. Becher, J. Eschner, F. Schmidt-Kaler, I.L. Chuangy, and R. Blatt, *Quantum information processing and cavity QED experiments with trapped $Ca^+$ ions*, in Atomic Physics **18** (Proceedings of the ICAP 2002).
http://heart-c704.uibk.ac.at/publications/papers/icap02_gulde.pdf

[40] J. Eschner, C. Raab, A. Mundt, A. Keuter, C. Becher, F. Schmidt-Kaler, and R. Blatt, Fortschritte der Physik **51**, 359 (2003).
http://heart-c704.uibk.ac.at/publications/papers/fortschritte03_eschner.pdf

[41] J.C. Bergquist, S.R. Jefferts, and D.J. Wineland, *Time Measurement at the Millennium*, Physics Today **54**, 1402 (2001).
http://tf.nist.gov/general/pdf/1402.pdf
http://tf.nist.gov/cesium/fountain.htm

[42] C. Monroe, D.M. Meekhof, B.E. King, S.R. Jefferts, W.M. Itano, D.J. Wineland, and P. Gould, Physical Review Letters **75**, 4011 (1995).
http://tf.nist.gov/general/pdf/1100.pdf

[43] A.E. Leanhardt, T.A. Pasquini, M. Saba, A. Schirotzek, Y. Shin, D. Kielpinski, D.E. Pritchard, and W. Ketterle, Science (Reports) **301**, 1513 (2003).

[44] J.J. McClelland and M.R. Scheinfein, J. Opt. Soc. Am. B **8**(9), 1974 (1991).
http://physics.nist.gov/Divisions/Div841/Gp3/Pubs/pdf/epg574.pdf

[45] E.C. Harvey, T.R. Mackin, B.C. Dempster, R.E. Scholten, *Micro-optical structures for atom lithography studies*, Proc. SPIE **3892**, 266 (1999).
http://optics.ph.unimelb.edu.au/~mackin/research/SPIE_paper_final.doc
S.M. Iftiquar, J. Opt. B: Quantum Semiclass. Opt. **5**, 40 (2003);
http://xxx.lanl.gov/abs/physics/0306021

[46] D.M. Eigler, C.P. Lutz, and W.E. Rudge, Nature **352**, 600 (1991); J.A. Stroscio and D.M. Eigler, Science **254**, 1319 (1991).

[47] G. Meyer, J. Repp, S. Zöphel, K-F Braun, S.W. Hla, S. Fölsch, L. Bartels, F. Moresco, and K.H. Rieder, Single Molecule **1**, 79 (2000).
http://www.physik.fu-berlin.de/~grill/manipulation/manipulationatoms/
http://www.physik.fu-berlin.de/~ag-rieder/pr4eo/manipulation/process.html

[48] J.Y. Park, R.J. Phaneuf, and E.D. Williams, Surface Sci. **470**, L69 (2000).
http://www.phy.ohiou.edu/~hla/HLA1999-1.pdf

[49] T.T. Tsong and C.S Chang, Jpn. J. Appl. Phys. **34**, 3309 (1995).

[50] L.S. Hordon, B.B. Boyer, and R.F.W. Pease, J. Vac. Sci. Technology B **13**, 826 (1995); http://scitation.aip.org/jvstb/

___

**FIGURES**

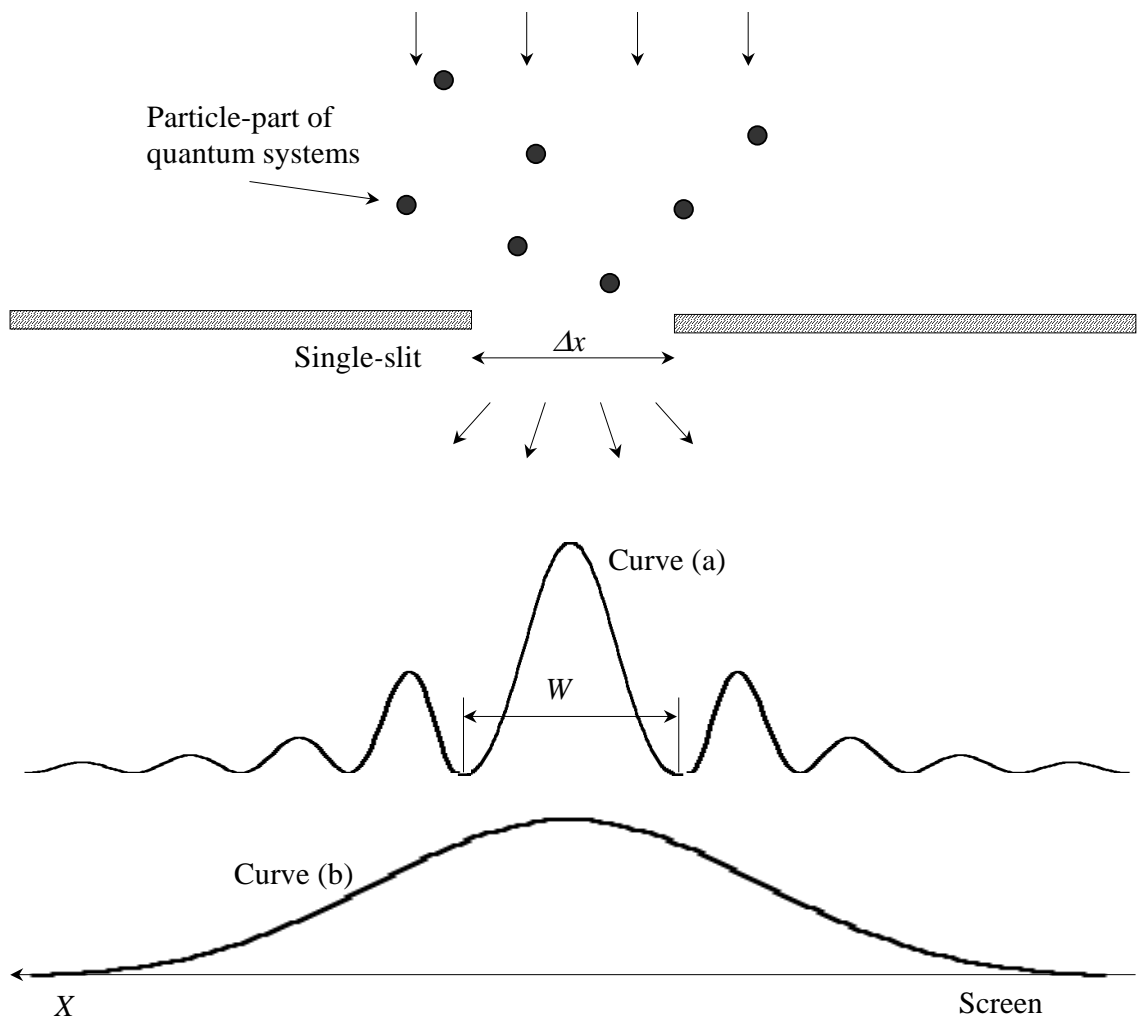

Figure 1. Single-slit diffraction experiment. Curve (a) is the usual diffraction density curve that can be observed on the screen in case of an incident beam that is at least as wide as $\Delta x$. Curve (b) represents the dispersion curve which is equivalent to curve (a) but in absence of any interference fringes.

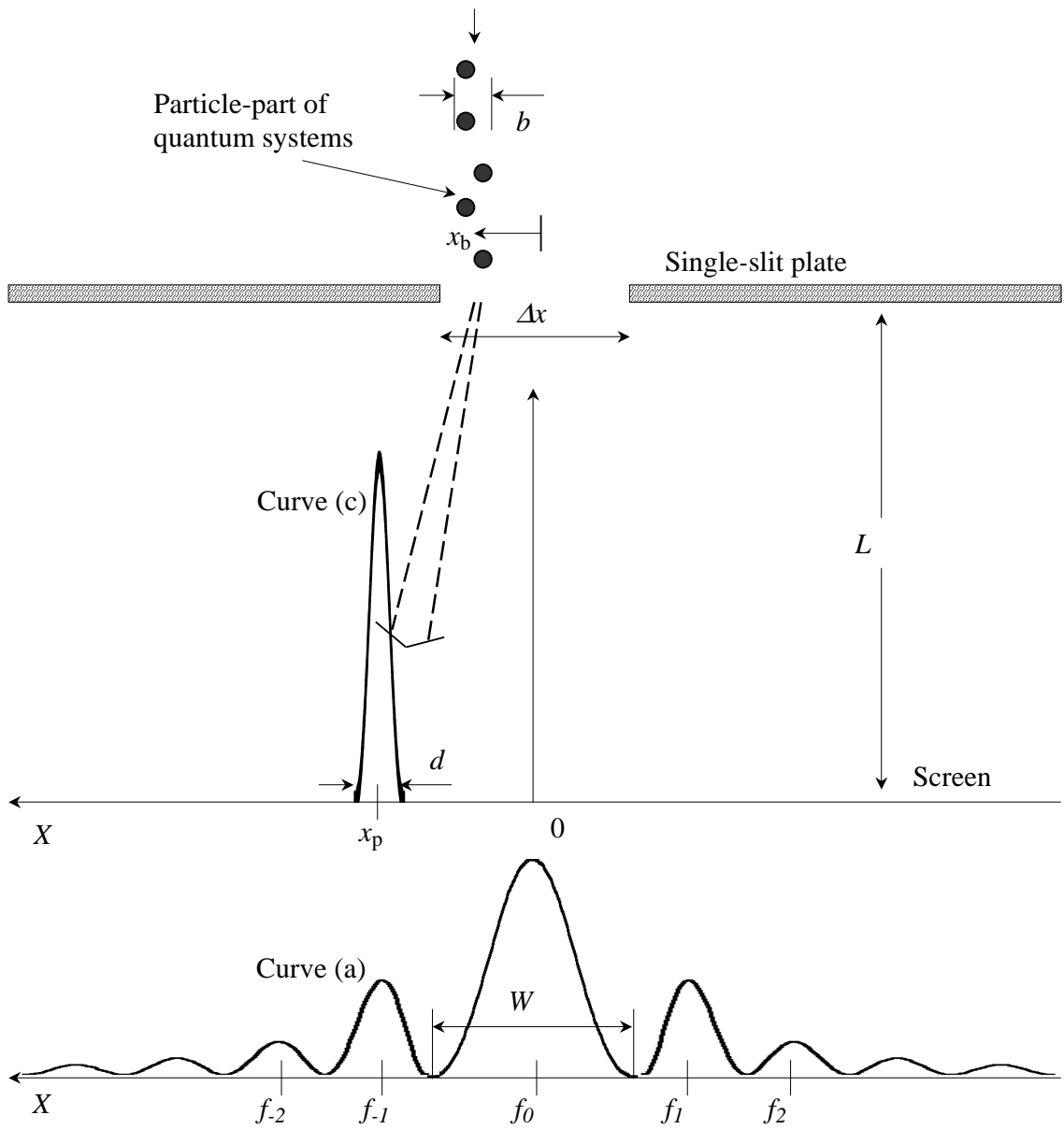

Figure 2. Proposed single-slit experiment with a fine beam input. Curve (a) is the usual diffraction curve obtained with the experimental setup and conditions of Fig. 1. Curve (c) is the expected diffraction curve that represents a narrow line on the screen.

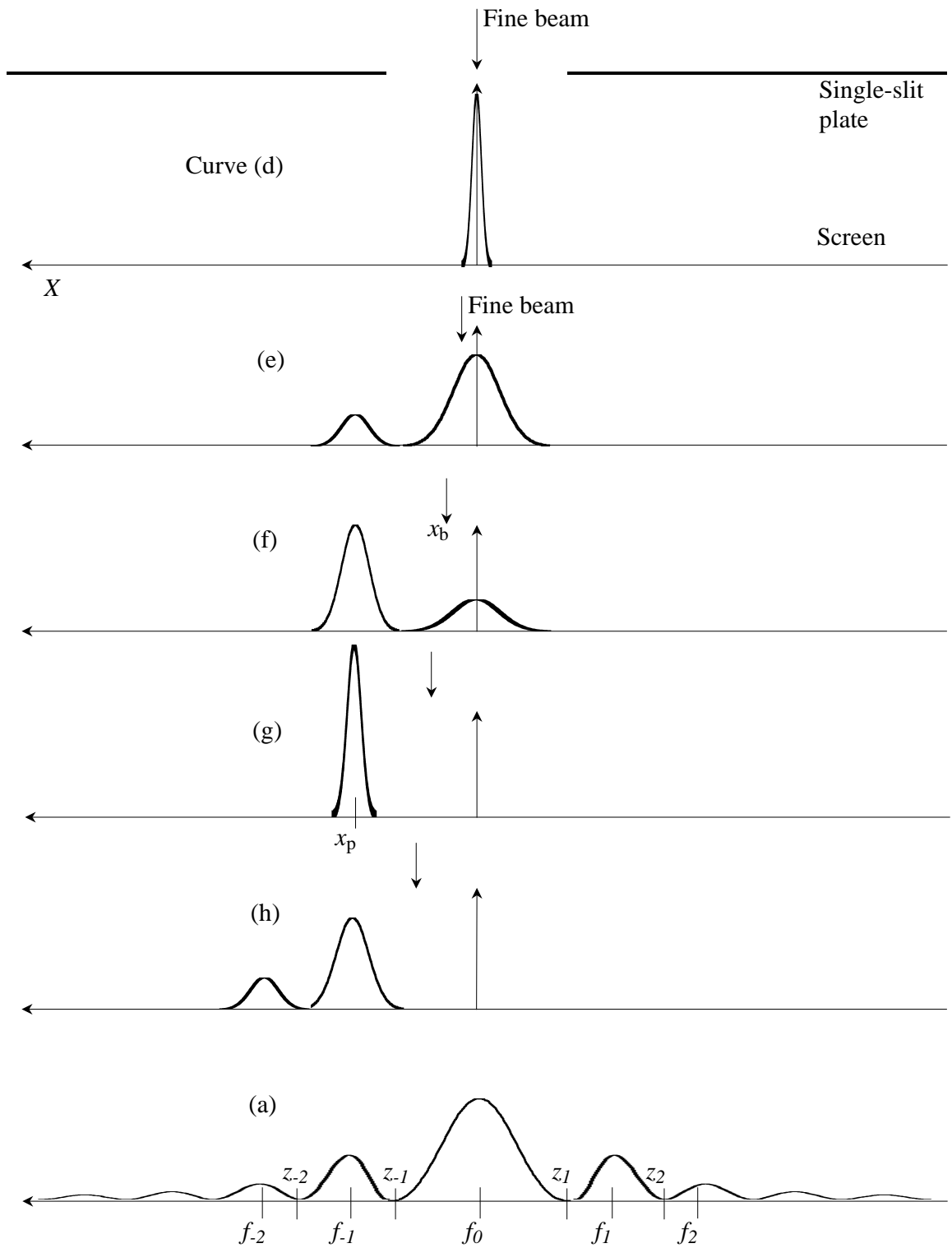

Figure 3. Curve (a): experimental setup and conditions of Fig. 1. Curves (d-h) is a sequence of expected diffraction curves as a function of $x_b$ for the case of a fine beam.